\documentclass[%
 reprint,
superscriptaddress,
 amsmath,amssymb,
 aps,
prb,
]{revtex4-1}

\usepackage{graphicx}
\usepackage{dcolumn}
\usepackage[colorlinks=true,linkcolor=blue]{hyperref}%
\usepackage{algorithm}
\usepackage{algpseudocode}
\usepackage{pifont}
\usepackage{hyperref}
\usepackage{natbib}

\expandafter\ifx\csname package@font\endcsname\relax\else
 \expandafter\expandafter
 \expandafter\usepackage
 \expandafter\expandafter
 \expandafter{\csname package@font\endcsname}%
\fi
\begin{document}

\preprint{APS/123-QED}

\title{A generalized Poisson and Poisson-Boltzmann solver\\ 
for electrostatic environments}

\author{G. Fisicaro}
\email{giuseppe.fisicaro@unibas.ch}
\affiliation{Department of Physics, University of Basel, Klingelbergstrasse 82, 4056 Basel, Switzerland}%

\author{L. Genovese}%
\affiliation{Laboratoire de simulation atomistique (L\_Sim), SP2M, INAC, CEA-UJF, Grenoble, F-38054, France}%

\author{O. Andreussi}%
\affiliation{Institute of Computational Science, Universita' della Svizzera Italiana,Via Giuseppe Buffi 13, CH-6904 Lugano}%
\affiliation{Theory and Simulations of Materials (THEOS) and National Centre for Computational Design and Discovery of Novel Materials (MARVEL), \'Ecole Polytechnique F\'ed\'erale de Lausanne, Station 12, CH-1015 Lausanne, Switzerland}%

\author{N. Marzari}%
\affiliation{Theory and Simulations of Materials (THEOS) and National Centre for Computational Design and Discovery of Novel Materials (MARVEL), \'Ecole Polytechnique F\'ed\'erale de Lausanne, Station 12, CH-1015 Lausanne, Switzerland}%

\author{S. Goedecker}%
\affiliation{Department of Physics, University of Basel, Klingelbergstrasse 82, 4056 Basel, Switzerland}%

\date{\today}

\begin{abstract}
The computational study of chemical reactions in complex, wet environments is critical for
applications in many fields. It is often essential to study chemical reactions in the presence
of applied electrochemical potentials, taking into account the non-trivial electrostatic screening
coming from the solvent and the electrolytes. As a consequence the electrostatic
potential has to be found by solving the generalized Poisson and
the Poisson-Boltzmann equation for neutral and ionic solutions, respectively.
In the present work solvers for both problems have been developed. 
A preconditioned conjugate gradient method has been implemented to the generalized Poisson
equation and the linear regime of the Poisson-Boltzmann, allowing to solve iteratively
the minimization problem with some ten iterations of a ordinary Poisson equation solver.  
In addition, a self-consistent procedure enables us to solve the non-linear Poisson-Boltzmann
problem. 
Both solvers exhibit very high accuracy and parallel efficiency,
and allow for the treatment of different boundary conditions, as for example surface systems.
The solver has been integrated into the BigDFT and Quantum-ESPRESSO electronic-structure packages
and will be released as an independent program, suitable for integration in other codes.
\end{abstract}

\pacs{Valid PACS appear here}
\maketitle

%
%
%
%
%
%
\section{Introduction}

Many important chemical processes take place in solution both in the context of 
basic and industrial research.
The computational study of such chemical reactions in wet environments
is therefore of cross-disciplinary interest to physics, chemistry, materials science,
chemical engineering, and biology\cite{Dabo_book2010}. Computational studies can complement  
these
investigations by giving insight into new processes and
materials as well as reducing development times and production costs.
Solar-energy harvesting in a dye-sensitized
cell or electro-catalytic water splitting are two simple examples of relevance for
applications in the energy and environment context.

Molecular properties in the presence
of a solution are often very different compared to pure 
\emph{in vacuum}\cite{Reichardt_Book2003} conditions making vacuum-like \emph{ab initio}
calculations an inappropriate approach for such problems. An inclusion of the solute-solvent
interaction in \emph{ab initio} simulations is thus mandatory.
On the atomistic scale, an explicit inclusion of all solvent molecules
in the simulation should be in principle the natural way to account for solvent effects.
Due to the very large number of water and possibly other molecules required, 
this approach enormously increases the computational cost and limits
at the same time the size of the system contained in the explicit
dielectric medium\cite{White_JCP2000}.
The study of the solute-solvent interaction
at length scales larger than the molecular sizes would become virtually impossible.
Investigations like structure predictions 
\cite{Goedecker_JCP2004} or reaction path
determination\cite{Schaefer_JCP2014} would become unaffordable in such a purely 
atomistic approach.
Moreover, fully atomistic simulations of solvation effects would need to deal with
the extensive sampling, required to characterize liquid configurations, and with
the well-known limitations that current state-of-the-art \emph{ab initio} methods present
in describing liquid water, in particular regarding its structural and dielectric properties.

An implicit inclusion of the solute-solvent interactions could solve
these issues. Starting from the earliest work of Onsager\cite{Onsager_JACS1936},
the quantum chemistry community investigated implicit solvation
models\cite{Tomasi_CR1994,Tomasi_CR2005,Cramer_CR1999} extensively. In these 
models the solvent
is introduced as a continuous homogeneous and isotropic medium fully described
by a dielectric function.
Among them, the polarizable continuum model (PCM) developed by Tomasi and
co-workers\cite{Tomasi_CR1994,Tomasi_CR2005} is one of the most popular.
In this approach a dielectric cavity
surrounding the atomistic system is introduced where the permittivity takes
on the value of one in regions occupied by atoms and some different value characteristic for the dielectric solvent medium considered outside. 

Density functional theory (DFT) is a widely used  method
to investigate material properties at the atomistic scale.
In such \emph{ab initio} calculations the electronic-structure problem
is solved by minimizing the total energy of the system which is a functional
of the electronic density

\begin{equation}
\label{ene}
E[\rho] = T[\rho] + \int v(\textbf{r}) \rho(\textbf{r}) \mathrm{d}\textbf{r} + \frac{1}{2} \int \rho(\textbf{r}) \phi[\rho] \mathrm{d}\textbf{r} + E_{xc}[\rho] ,
\end{equation}

where the four terms on the right side of Eq. (\ref{ene}) are, respectively, the standard kinetic energy,
the interaction energy with an external potential, the electrostatic and the exchange-correlation energy.
For gas-phase molecular simulations the potential $\phi(\textbf{r})$
generated by a given charge density $\rho(\textbf{r})$ is given by the solution of the standard
Poisson (SPe) equation

\begin{equation}
\label{SPe}
\nabla^2 \phi(\textbf{r}) = -4 \pi \rho(\textbf{r}) .
\end{equation}

An implicit inclusion of the solvent can be obtained by introducing
a continuum dielectric cavity by means of a
dielectric distribution $\epsilon( \textbf{r})$. Then the potential is given by the solution
of the generalized Poisson equation (GPe)

\begin{equation}
\label{GPe}
\nabla \cdot \epsilon( \textbf{r}) \nabla \phi(\textbf{r}) = -4 \pi \rho(\textbf{r}) .
\end{equation}

If the system is surrounded by an ionic solution, an extra-term has to be added on the right
side of Eq. (\ref{GPe}). It accounts for the ionic distribution in the liquid and
depends on the local electrostatic potential $\phi(\textbf{r})$. The resulting
non-linear differential equation would therefore become
\begin{equation}
\label{PBe}
\nabla \cdot \epsilon( \textbf{r}) \nabla \phi(\textbf{r}) = -4 \pi \left( \rho(\textbf{r}) + \rho^{ions}[\phi](\textbf{r}) \right) ,
\end{equation}
where $\rho^{ions}(\textbf{r})$ is the local ionic concentration of the ions
in the dielectric solvent,
written as a sum of concentration contributions $c_i$ of ions of type $i \in \{ 1,2,...,m \} $ and valence $Z_i$,
which in turn are $\phi$-dependent functionals:
\begin{equation}
\label{rhoions}
\rho^{ions}[\phi](\textbf{r}) = e N_{\text{A}} \sum_{i=1}^{m} 
Z_i c_{i}[\phi](\textbf{r}) \, ,
\end{equation}
where $e$ is the elementary charge and $N_{\text{A}}$ the Avogadro's number.
The most common expression for the $c_i[\phi]$ functional gives rise to the well-known Poisson-Boltzmann equation (PBe).

Whereas several approaches and solvers exist for the SPe\cite{Greengard_JCP1987,Strain_Science1996,Genovese_JCP2006,Genovese_JCP2007,Cerioni2013},
efficient and accurate solvers are still missing for the GPe and PBe cases.
Some of these solvers handle \emph{sparse} matrices obtained by low-order finite difference
discretization of Eq. (\ref{GPe}), many of them keeping constant the permittivity
in both inner and external regions of the dielectric cavity and then solving
a standard Poisson problem.
Furthermore, simpler and linearized forms of the Poisson-Boltzmann equation 
are usually considered, neglecting steric effects and overestimating ionic concentrations
close to highly charged surfaces and for multivalent ions.

Fast and accurate GPe and PBe solvers which accurately works for
a continuously varying dielectric function could therefore play a key role in the extension of 
vacuum-based atomistic packages and allowing for quantum simulations in the presence of water,
dissolved species, electrolytes, and non-aqueous solvents.

In the present paper we present a minimization technique which
solves the generalized Poisson problem in some ten iterations of a SPe solver. 
In combination with a self-consistent
procedure it enables us to solve the non-linear Poisson-Boltzmann problem
in a formulation which include ionic steric effects.
The implemented algorithms take advantage of the chosen preconditioner
for the minimization procedure.
The accuracy of both GPe and PBe solvers has been tested for cases where an analytic solution is available.
Two different methods have been implemented to describe the solvent
surrounding the atomistic system. We prove the effectiveness of our method in practical DFT calculations of electrostatic solvation energies of various test systems.

\section{\label{GPequation} Generalized Poisson equation}
As discussed above, the dielectric medium is described by means of a position-dependent dielectric
distribution $\epsilon( \textbf{r})$.
In order to solve numerically Eq. (\ref{GPe}), both the potential $\phi(\textbf{r})$, the charge density
$\rho(\textbf{r})$ and the dielectric function $\epsilon(\textbf{r})$ are generally discretized on a finite grid.
In principle also the generalized Poisson operator

\begin{equation}
\label{GPop}
\mathcal{A} = \nabla \cdot \epsilon( \textbf{r}) \nabla
\end{equation}

should be discretized on the same mesh.
It will be shown that depending on the adopted strategy to solve
numerically Eq. (\ref{GPe}), the discretization of the differential operator $\mathcal{A}$ can be avoided
in exchange of a iterative procedure based on a SPe solver.

An alternative would be to solve the GPe iteratively as suggested in Ref. [\citenum{Andreussi_JCP2012}].
In this approach the polarization field introduced by the spatially varying dielectric function
is added as a source term to the charge density of the ordinary Poisson equation
and the Poisson equation is solved repeatedly until self-consistency between the potential
and the polarization charge density induced by it is reached.
Our approach completes and simplifies this treatment, reducing
considerably the number of SPe iterations, thereby presenting a
robust and powerful iterative solver.

Considering that reliable convergence can be an issue in mixing schemes, an alternative approach based on a minimization procedure
is desireable. Such an approach can be based on an action integral
whose Euler Lagrange equation is the GPe [Eq. (\ref{GPe})]:
\begin{equation}
\label{AIntegral}
\mathcal{I} = \int \left[ \frac{1}{2} \nabla \phi(\textbf{r}) \epsilon(\textbf{r}) \nabla \phi(\textbf{r})) -4 \pi \rho(\textbf{r}) \phi(\textbf{r}) \right]\mathrm{d}\textbf{r} \;.
\end{equation}
Any numerical minimization scheme can then be applied to solve the electrostatic problem.

In Sec. \ref{PCGprocedure} our strategy to solve Eq. (\ref{GPe}),
based on a preconditioned conjugate gradient (PCG) algorithm, will be presented. The preconditioner
exactly represents the operator in the limit of a slowly varying dielectric constant
and is based on the standard Poisson solver of BigDFT.

The possibility of solving the generalized Poisson equation under various boundary conditions (BC) will be
very important. In our approach the boundary conditions
enter in a straightforward way by means of the preconditioner,
i.e through the solution of the SPe.

\subsection{\label{SCprocedure} Self-consistent iterative procedure}

A strategy to solve the GPe for a given charge density $\rho(\textbf{r})$ is by means of a self-consistent (SC) iterative procedure \cite{Andreussi_JCP2012}.
Applying simple algebraic manipulations, Eq. (\ref{GPe}) can be rewritten as

\begin{equation}
\label{remap}
\begin{split}
\nabla^2 \phi(\textbf{r}) & = -4 \pi \left[ \frac{\rho(\textbf{r})}{\epsilon(\textbf{r})}+ \rho^{iter}(\textbf{r}) \right]\\
& = -4 \pi \left[ \rho(\textbf{r})+ \rho^{pol}(\textbf{r}) \right]\;,
\end{split}
\end{equation}

where $\rho^{pol}(\textbf{r})$ is the polarization charge density.
In this approach an extra-term $\rho^{iter}(\textbf{r})$
\begin{equation}
\label{rhoiter}
\rho^{iter}(\textbf{r}) =\frac{1}{4 \pi} \nabla \ln \epsilon(\textbf{r}) \cdot \nabla \phi(\textbf{r})
\end{equation}
induced by the spatially-varying dielectric function $\epsilon(\textbf{r})$
is added as a source to the charge density of the ordinary Poisson equation.
Hence the GPe can be solved by a self-consistent loop on the potential
$\phi(\textbf{r})$, obtained by a SPe solver onto the second member of Eq.~\eqref{remap}. The residual $r_k$, quantifying the convergence, is the difference between the extra terms of Eq. (\ref{rhoiter}) between subsequent iterations.
Algorithm \ref{SC} describes the procedure. In order to stabilize the iterative method,
a linear mixing of the extra-term $\rho^{iter}(\textbf{r})$ at steps $k$th and $(k+1)$th
has been introduced tuned by means of the mixing parameter $\eta$.

\begin{figure}
\begin{algorithm}[H]
\linespread{1.5}\selectfont
\caption{Self-consistent iterative procedure (SC)}
\label{SC}
\begin{algorithmic}[1]
\State $\rho^{iter}_{0}=0$
\For{$k = 0,1,\cdots$}
\State $\rho^{tot}_{k} = \rho/\epsilon + \rho^{iter}_{k}$
\State solve $\nabla^2 \phi_{k} = -4 \pi \rho^{tot}_{k}$
\State $\rho^{iter}_{k+1} =\frac{1}{4 \pi} \nabla \ln \epsilon \cdot \nabla \phi_{k}$
\State $\rho^{iter}_{k+1} = \eta \rho^{iter}_{k+1} + (1-\eta)\rho^{iter}_{k}$
\State $r_{k+1} = \rho^{iter}_{k+1} - \rho^{iter}_{k}$
\EndFor
\end{algorithmic}
\end{algorithm}
\end{figure}

The polarization charge induced in the dielectric medium can be easily
related to the extra-term of Eq. (\ref{rhoiter}):
\begin{equation}\label{rhopol}
\rho^{pol}(\textbf{r}) = \rho^{iter}(\textbf{r}) + \frac{\epsilon(\textbf{r})-1}{\epsilon(\textbf{r})} \rho(\textbf{r}) .
\end{equation}

This charge represents the response of the surrounding implicit dielectric. It lies in the transition
region between the inner and outside part of the cavity and stabilizes the solute density
enveloped by the solvent.

\subsection{\label{PCGprocedure} Preconditioned conjugate gradient}

Although a reasonably small number of iterations can be obtained in a
self-consistent scheme, minimization techniques can produce more efficient methods
to handle Eq.(\ref{GPe}) if sophisticated schemes are utilized.
In addition, the formulation as a minimization problem would allow to better control
the convergence behavior.

Solving this equation with a preconditioned steepest descent (PSD) method is essentially identical
to the self-consistency approach of Sec. \ref{SCprocedure}, once a standard
Poisson solver is taken as preconditioner.
In particular, a good preconditioner for a PSD minimization, which inverse applied to a
residual vector $r_k$ provides the preconditioned residual $v_k$, is as follows:
\begin{equation}
\label{PrePSD}
\mathcal{P^{\textit{SD}}} v_k(\textbf{r}) =  \epsilon(\textbf{r}) \nabla^{2} v_k(\textbf{r}) = -4 \pi r_k(\textbf{r}) .
\end{equation}

Being the Laplacian of $v_{k}(\textbf{r})$ related to the residual
vector $r_k$ by means of Eq. (\ref{PrePSD}), the generalized Poisson
operator becomes

\begin{equation}
\begin{split}
\mathcal{A} v_k(\textbf{r}) & = \nabla \cdot \epsilon( \textbf{r}) \nabla v_k(\textbf{r}) \\
& = \nabla \epsilon( \textbf{r}) \cdot \nabla v_k(\textbf{r}) - 4 \pi r_k(\textbf{r}) .
\end{split}
\end{equation}

Such a PSD approach can be described by Algorithm \ref{PCG} with $\beta_k =0$.
Fixing $\alpha_k=1$ it corresponds to the previously described
self-consistent approach.

In a PSD scheme the number of iterations $l$ needed for convergence is proportional to the condition
number $\kappa$ (i.e. the ratio between the largest and the smallest eigenvalue of the
product operator $\mathcal{P}^{-1}\mathcal{A}$).
Minimization methods with a faster convergence rate than the preconditioned steepest descent
algorithm can significantly improve the convergence speed.

We use a preconditioned conjugate gradient scheme, where $l \propto \sqrt{\kappa}$.
In such minimization procedure a good preconditioner can lower
$\kappa$ and, therefore, the overall number of iterations.
Algorithm \ref{PCG} describes the implemented PCG procedure to compute the electrostatic
potential $\phi(\textbf{r})$ starting from a given charge density $\rho(\textbf{r})$.
The minimization procedure starts from an initial gradient $r_0$
computed on an input guess $\phi_0$.
$\mathcal{P}$ is the preconditioner which inverse has to be applied to the residual vector
$r_k$ returning the preconditioned residual $v_k$, and, finally, $\phi_k$ is the solution
of Eq. (\ref{GPe}).
The convergence criterion is imposed on the Euclidean
norm of the residual vector $r_k$.

\begin{figure}
\begin{algorithm}[H]
\linespread{1.5}\selectfont
\caption{Preconditioned conjugate gradient}
\label{PCG}
\begin{algorithmic}[1]
\State $r_0 = -4 \pi \rho - \mathcal{A} \phi_0$, $p_{-1} = 0$
\For{$k = 0,1,\cdots$}
\State $ v_k = \mathcal{P}^{-1} r_k$
\State $ p_k = v_k + \beta_k p_{k-1} $ (where $\beta_k = \frac{(v_k,r_k)}{(v_{k-1},r_{k-1})}, k \neq 0 $ )
\State $ \alpha_k = \frac{(v_k,r_k)}{(p_k,Ap_k)} $
\State $ \phi_{k+1} = \phi_k + \alpha_k p_k $
\State $ r_{k+1} = r_k - \alpha_k \mathcal A p_k $
\EndFor
\end{algorithmic}
\end{algorithm}
\end{figure}

Both the performance and accuracy in a PCG scheme critically depend on the preconditioner chosen.
We implemented a preconditioner based on the solution of the standard Poisson equation
(namely on a standard Posson solver).
Once a residual vector $r_k$ is given at the step 3 of Algorithm \ref{PCG}, we define
a preconditioned residual from the following
equation:
\begin{equation}\label{PCGdef}
\mathcal{P^{\textit{CG}}} v_k(\textbf{r}) = \sqrt{\epsilon(\textbf{r})} \nabla^{2} [ v_k(\textbf{r}) \sqrt{\epsilon(\textbf{r})}] = -4 \pi r_k(\textbf{r}) .
\end{equation}

This equation has to be solved with respect $v_k(\textbf{r})$ once $r_k(\textbf{r})$ is given.

In addition to speeding up the PCG procedure, the preconditioner defined by Eq.~\eqref{PCGdef}
retains a further feature which guarantees accuracy and fast performance for the whole
electrostatic solver.
In step 7 of Algorithm \ref{PCG} we have to apply the generalized Poisson operator $\mathcal{A}$ to
the preconditioned residue $p_k$, which means, thanks to step 3, applying it to $v_k$.
Using a change of variable $v^{\prime}_k(\textbf{r}) = \sqrt{\epsilon(\textbf{r})} v_k(\textbf{r})$,
the GPe becomes
\begin{equation}
\label{Op_makeup}
\nabla \cdot \epsilon( \textbf{r}) \nabla v_k(\textbf{r}) = \sqrt{\epsilon(\textbf{r})} \nabla^{2} v^{\prime}_k(\textbf{r}) -v^{\prime}_k(\textbf{r}) \nabla^2 \sqrt{\epsilon(\textbf{r})} .
\end{equation}

Now simple algebraic manipulations and Eq. (\ref{PCGdef}) allow to rewrite the
generalized Poisson operator $\mathcal{A}$ as 
\begin{align}
\mathcal{A} v_k(\textbf{r}) & = \nabla \cdot \epsilon( \textbf{r}) \nabla v_k(\textbf{r}) \\
& = -v_k(\textbf{r}) q(\textbf{r}) -4 \pi r_k(\textbf{r}) ,
\label{corr}
\end{align}

where $q(\textbf{r}) = \sqrt{\epsilon(\textbf{r})} \nabla^2 \sqrt{\epsilon(\textbf{r})}$ is
calculated once at the beginning of the PCG procedure and kept fixed for the whole minimization loop.
Therefore thanks to the chosen preconditioner, the action of the operator $\mathcal{A}$ can be
simplified to a simple multiplication between
the potential $v_k(\textbf{r})$ and a constant vector $q(\textbf{r})$ related to the
spatially-varying dielectric function $\epsilon(\textbf{r})$.
This feature, which provides the exact operator output, makes our PCG procedure robust
and fast, avoiding any finite difference differentiation.
Furthermore, reducing the PCG algorithm to simple vector operations, makes its parallelization
straightforward, delegating it to the chosen SPe solver.
A similar discussion holds for the boundary conditions, which enter in a natural way by means of
the preconditioner, i.e through the solution of the ordinary Poisson equation,
both in the SC and PCG algorithm.

\subsection{\label{Nume} Numerical results}

Both the self-consistent iterative procedure (Algorithm \ref{SC}) and the preconditioned
conjugate gradient minimization scheme (Algorithm \ref{PCG}) have been implemented and tested.
As SPe solver (step 4 of Algorithms \ref{SC} and step 3 of Algorithm
\ref{PCG}), we used the Interpolating Scaling Function (ISF) Poisson Solver, 
allowing to obtain highly accurate electrostatic potentials for free, wire, surface, and periodic
boundary conditions at the cost of $O(N\log(N))$ operations, where $N$ is the number of discretization points
(see Ref. [\citenum{Cerioni2013}]).

\begin{figure}
\includegraphics[width=0.5\textwidth]{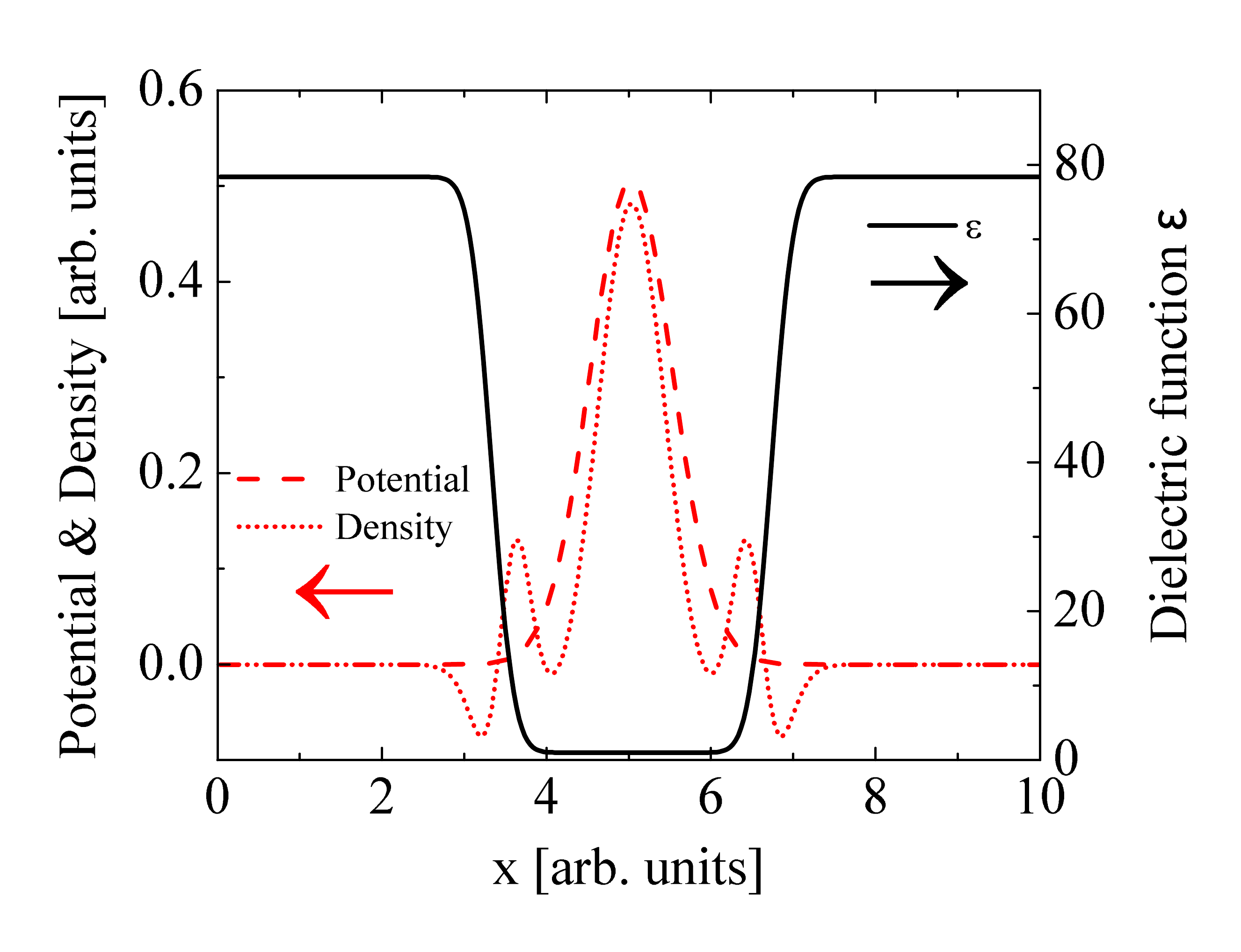}
\caption{\label{AnFunc}Analytical three dimensional functions used as benchmark fields for
both SC and PCG solvers along a particular direction passing though the box center and parallel to
the y axis. Red dash line: potential $\phi(\textbf{r})$; red dot line: charge density $\rho(\textbf{r})$;
black solid line: dielectric function $\epsilon(\textbf{r})$.}
\end{figure}

To test both solvers analytic three dimensional functions have
been used. An orthorhombic grid of uniform mesh spacing
$h_{\textnormal{grid}}$ and $(n_x,n_y,n_z)$ points in each
directions has been used.
Fig. \ref{AnFunc} shows plots of these benchmark fields along a particular direction passing through the box
center and parallel to the y axis.
All functions depend on the radial distance $r$
from the center of the simulation domain.

\begin{figure*}
\includegraphics[width=1.0\textwidth]{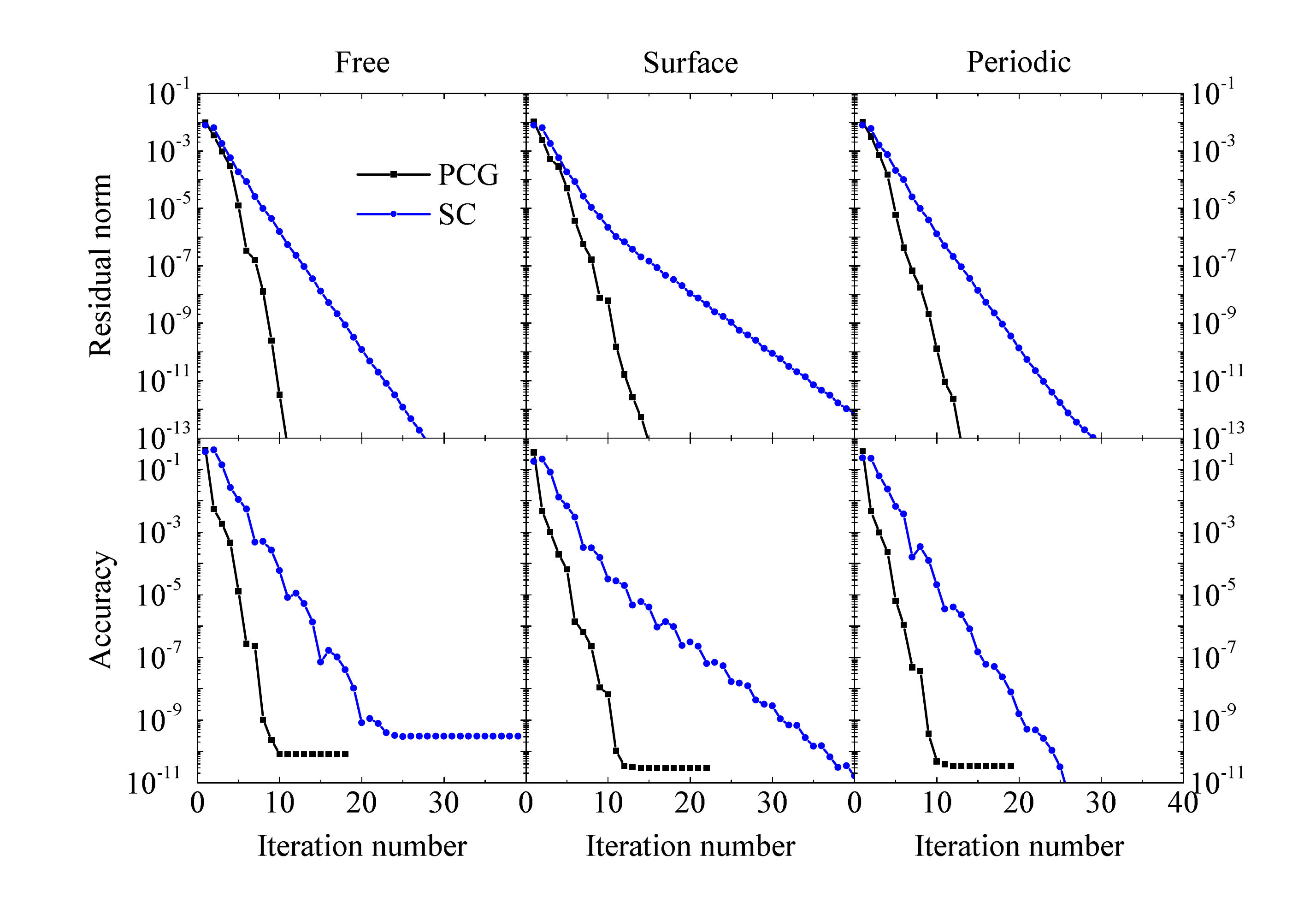}
\caption{\label{NormrAcc}Euclidean norm of the residual vector $r_k$ (top graphs) and accuracy of the
numerical solution (lower graphs) for SC (blue circles) and PCG (black squares)
solvers with free, surface and period boundary conditions.}
\end{figure*}

A normalized Gaussian function has been chosen for the electrostatic potential $\phi(\textbf{r})$
(red dash line in Fig.~\ref{AnFunc}) and the charge density $\rho(\textbf{r})$ has been derived from the chosen
potential and dielectric function, applying the generalized Poisson differential operator
of Eq. (\ref{GPop}) (red dot line). In order to reproduce the dielectric environment in electrostatic
problems typically used for a solute system embedded in a solvent (i.e. a cavity
where the majority of the atomic charge density is confined), the error
function $ 1+ (\epsilon_0-1) h(d_0,\Delta;r)$ has been chosen to represent the spatially varying electric constant $\epsilon({\bf r})$ (solid black line in Fig.~\ref{AnFunc}), where
\begin{equation}\label{hfunc}
h(d_0,\Delta;r)= \frac{1}{2} \left[ 1 + \mathrm{erf} \left( \frac{r-d_{0}}{\Delta} \right) \right]
.
\end{equation}
Here $\Delta$ is a parameter which controls the transition region ($\approx 4 \Delta$ wide)
between the inner part within the cavity of radius $d_0$ and the external part, and $\epsilon_0$ is the dielectric constant of the surrounding medium.
These benchmark functions have been used for both free, surface and periodic boundary conditions.

To implement the nabla differential operator needed for Algorithm \ref{SC}, central,
forward and backward finite difference filters of order 16 have been used,
which match the accuracy of the underlying SPe solver.
We remark that the use of these finite difference filters is not needed
for Algorithm \ref{PCG} as soon as the correction term of Eq.~\eqref{corr} is pre-calculated.

Fig. \ref{NormrAcc} shows solver performances. The top graphs report the residual norm
as function of the iteration number. The residual
norm is defined has the Euclidean norm of the residual vector $r_k$.
Graphs in lower panel present the output accuracy as a function of the iteration number. The accuracy in the whole paper is defined
as the maximum value of the difference between the final numerical solution and
the analytical potential. 
In this test case a cubic box of length 10 [arb. units] has been chosen with $n_x=n_y=n_z=300$.
The Gaussian variance for the potential $\phi(\textbf{r})$ was $\sigma=0.5$, and the parameters of
$\epsilon({\bf r})$ were $d_0=1.7$, $\Delta=0.3$ and $\epsilon_0=78.36$ (all in [arb. units]).
The mixing parameter in step 6 of Algorithm \ref{SC} has been fixed to be $\eta=0.6$,
resulting in a robust convergence for all cases. Lower values slow down the convergence for the chosen test functions.

The PCG solver (black squares) exhibits a faster convergence with respect to the SC one (blue circles),
reaching an accuracy of $\sim 10^{-10}$ with some ten iterations. Furthermore its behavior does not change
with the boundary conditions as is the case with the SC algorithm.
It is worth remembering that each PCG iteration involves only a single solution of the ordinary Poisson equation
and as well as of fully parallelizable vector operations.
If an accuracy of $\sim 10^{-4} - 10^{-5}$ is enough, then some five iterations solve the electrostatic problem.
These features make the developed PCG algorithm together with the chosen preconditioner of Eqs. (\ref{PCGdef})
very efficient for atomistic calculations where the generalized Poisson equation needs to be solved repeatedly.
Performances of the implemented PCG procedure are also higher than
multigrid approaches to solve the GPe, where a number of iterations
between 17 and 25 are needed to reach an accuracy of $\sim 10^{-8}$ \cite{Fattebert_IJQC2003}.

For the sake of completeness, the preconditioned steepest descent scheme (Algorithm \ref{PCG} with $\beta_k =0$)
has been tested with the preconditioner described by Eq. (\ref{PrePSD}). Residual norm convergence as well as
the accuracy of the solution (not reported) behaves like the self-consistent approach
(Algorithm \ref{SC}, blue circles of Fig. \ref{NormrAcc}).
An integration of DIIS in the PSD loop effectively lowered the iteration numbers, but didn't show better
performances with respect to the PCG approach.
 
\section{\label{PBequation} Poisson-Boltzmann equation}

The generalized Poisson equation so far discussed holds for a solute plunged in a neutral
solution where no mobile charges are present. In order to extend the library to ionic solutions,
effects of mobile ions have to be taken into account which, being free to move inside the dielectric medium,
modify the spatial distribution of charge and potential close to the interface and giving rise
to the well known double layer.

In general, mobile charges can be included in the electrostatic
problem by means of a continuum mean-field approach, assuming
point-like ions in thermodynamic equilibrium. Once the equilibrium is reached, ionic
concentrations explicitly depend on the local electrostatic potential $\phi(\textbf{r})$.
Following this idea, the potential $\phi(\textbf{r})$ generated by
a charge density $\rho(\textbf{r})$ placed in contact with a ionic solution
can be extracted solving the generalized Poisson equation (Eq.~\eqref{PBe}).
Several models have been proposed in the literature for the ionic bulk concentrations [see Eq.~\eqref{rhoions}].
Gouy\cite{Gouy_JPTA1910} and Chapman\cite{Chapman_PM1913} proposed a Boltzmann distribution 
\begin{equation}
\label{ionsconcPB}
c_{i}[\phi](\textbf{r}) = c_{i}^{\infty} \exp \left( -\frac{Z_{i} e \phi(\textbf{r})}{kT} \right) ,
\end{equation}
where $k$ is the Boltzmann constant and $T$ the absolute temperature of the solvent.
Combining Eqs. (\ref{PBe},\ref{rhoions},\ref{ionsconcPB}) the well-known Poisson-Boltzmann equation can be recovered.
It arises from the equilibrium between thermal and electric forces, which depend, respectively,
on the ionic concentration and electrostatic potential.
At equilibrium the total average force
must be zero and Eq. (\ref{ionsconcPB}) holds.
In the regions where the electrostatic energy is smaller then
$kT$, i.e. $Z_{i} e \phi(\textbf{r}) / kT \ll 1$, the
exponential of Eq. (\ref{ionsconcPB}) can be approximated with a
linear function of $\phi(\textbf{r})$, switching the non-linear problem
of Eq.~\eqref{PBe} into a linear one.

The Poisson-Boltzmann equation correctly predicts ionic profiles close at the solid-liquid interface
with ionic solutions. However it strongly overestimates ionic concentrations close to highly
charged surfaces or multivalent ions. In order to overcome these drawbacks, several models
have been proposed\cite{Stern_E1924}.
Finite ion size effects can be included in the model by introducing an additional internal force.
Using a Bikerman-type expression to model steric effects \cite{Bikerman_PM1941} and imposing
that at equilibrium the total average force (thermal, electric and steric) must go to zero,
a Langmuir-type equation for the ionic concentrations can be found

\begin{equation}
\label{ionsconcMPB}
c_{i}[\phi](\textbf{r}) = \frac{ \displaystyle c_{i}^{\infty} \exp \left( -\frac{Z_{i} e \phi(\textbf{r})}{kT} \right)}{ \displaystyle 1 + \sum_{j=1}^{m} \frac{c_{j}^{\infty}}{c_{j}^{\text{max}}} \left[ \exp \left( -\frac{Z_{j} e \phi(\textbf{r})}{kT} \right) + 1 \right] } .
\end{equation}

In Eq. (\ref{ionsconcMPB}) $c_{i}^{\text{max}}$ are the maximum local concentration
that a ionic species of effective radius $R_{i}$ can attain.
This last is given by

\begin{equation}
\label{maxconc}
c_{i}^{\text{max}} = \frac{ \displaystyle p}{ \displaystyle \frac{4}{3} \pi R_{i}^{3} N_{\text{A}} } ,
\end{equation}

where $p$ is the packing coefficient. The combination of Eqs. (\ref{PBe},\ref{rhoions},\ref{ionsconcMPB})
give rise to the so-called modified Poisson-Boltzmann (MPBe) equation\cite{Otani_PRB2006,Jinnouchi_PRB2008,Dabo_book2010,Dabo_PhDThesis2008}.
It accounts for finite ion size effects representing an extension of the PBe,
preventing ion concentrations to exceed $c_{i}^{\text{max}}$.
Note that if $c_{i}^{\text{max}} \rightarrow \infty$ $\forall i \in \{ 1,2,...,m \}$,
Eq. (\ref{ionsconcMPB}) is reduced to Eq. (\ref{ionsconcPB}) and the standard Poisson-Boltzmann
equation holds.

Both PBe and MPBe equations consider ions in solution as pointlike. A further extension can account for
finite size effects in an explicit way, describing ions by means of
insulating spheres plunged in the
dielectric solvent \cite{Garcia_L2011}.
This correction allows for a local modification of the solvent permittivity
as well as the inclusion of two new forces: one related to a non-uniform dielectric which
tend to move ions into high-permittivity regions; another due to the
interaction of the ion dipole and the non-uniform local electric field (dielectrophoretic force).

The Poisson-Boltzmann equation is more difficult to solve than the generalized Poisson equation
due to the fact that it leads to a non-linear problem.
If $Z_{i} e \phi(\textbf{r}) / kT \ll 1$, Eq. (\ref{ionsconcPB})
can be approximated by a linear function of $\phi(\textbf{r})$, and the
Poisson-Boltzmann problem of Eq.~\eqref{PBe} becomes linear.
In Sec. \ref{PCGprocedure} a particular preconditioner, i.e.
Eq.(\ref{PCGdef}), has been proposed
for the solution of the GPe. Furthermore the generalized
Poisson operator has a linear term with respect to
$\phi(\textbf{r})$ [Eq. (\ref{corr})].
Therefore Algorithm \ref{PCG} is expected to solve the linear
regime of the PBe, where the $\mathcal A$ operator is now given by
Eq. (\ref{corr}) with an additional linear term:
\begin{multline}
\mathcal{A} v_k(\textbf{r})  = \nabla \cdot \epsilon( \textbf{r}) \nabla v_k(\textbf{r}) + 4 \pi \rho^{ions}[\phi](\textbf{r}) \\
 = -v_k(\textbf{r})\left( q(\textbf{r})+\frac{4 \pi e^{2} N_{\text{A}}}{kT} \sum_{i=1}^{m} Z_{i}^{2} c_{i}^{\infty} \right) -4 \pi r_k(\textbf{r})\; ,
\label{corrPB}
\end{multline}

where $q(\textbf{r})$ has been defined in Sec. \ref{PCGprocedure}.
In the general non-linear case, at variance to the generalized Poisson equation discussed in the previous sections,
there is no general consensus on what flavor of the Poisson-Boltzmann equation is most
appropriate for various problems.
Therefore a (M)PBe solver should allow to solve various formulations of this equation.
Since it is unlikely that for all variants an action integral can be established
(allowing for a minimization scheme), a self-consistent approach should be the most
appropriate algorithm.

Such a procedure has been implemented and it is detailed in Algorithm \ref{SCPB}.
The ionic charge density $\rho^{ions}[\phi]$ is included as source term to the charge density
of the generalized Poisson equation which, in turn, is solved repeatedly until
self-consistency between the potential and the ionic charge induced by it is reached.
Starting with an initial input guess
$\rho^{ions}_{0}$ for the ionic charge density, on each self-consistent cycle a generalized Poisson
solver (Algorithm \ref{SC} or \ref{PCG}) is applied to numerically compute the electrostatic
potential $\phi(\textbf{r})$. Then the ionic charge is computed using the new potential
and mixed with the old density by means of a linear scheme tuned by the parameter $\eta$.

In order to speed up performances of the (M)PBe solver, an improved
version is reported in Algorithm \ref{SCPBimproved}. On each iteration 
$k$ the electrostatic problem at step 4 is solved only for the previous
residual vector $r^{\prime}_{k-1}$ once $\rho^{ions}_{k}$ has been
updated at step 3. Then the overall solution is given at step 5 as sum
over all potential corrections $\phi^{\prime}_{k}$.
This procedure substantially corresponds to the general self-consistent
approach of Algorithm \ref{SCPB}, now using on each step information
of the previous as input guess. It is worth noting that the improved
Algorithm \ref{SCPBimproved} can be coupled only with the faster
PCG solver (Algorithm \ref{PCG}) for the GPe.

\begin{figure}
\begin{algorithm}[H]
\linespread{1.5}\selectfont
\caption{Self-Consistent iterative procedure for the Poisson-Boltzmann equation}
\label{SCPB}
\begin{algorithmic}[1]
\State set $\rho^{ions}_{0}$
\For{$k = 0,1,...$}
\State $\rho^{tot}_{k} = \rho + \rho^{ions}_{k}$
\State solve $\nabla \cdot \epsilon \nabla \phi_{k} = -4 \pi \rho^{tot}_{k}$ (Algorithm \ref{SC} or \ref{PCG})
\State compute $\rho^{ions}_{k+1} = \rho^{ions}[\phi_{k}]$
\State $\rho^{ions}_{k+1} = \eta \rho^{ions}_{k+1} + (1-\eta)\rho^{ions}_{k}$
\State $r_{k+1} = \rho^{ions}_{k+1} - \rho^{ions}_{k}$
\EndFor
\end{algorithmic}
\end{algorithm}
\end{figure}

\begin{figure}
\begin{algorithm}[H]
\linespread{1.5}\selectfont
\caption{Improved Self-Consistent iterative procedure for the Poisson-Boltzmann equation}
\label{SCPBimproved}
\begin{algorithmic}[1]
\State set $\phi_{1}=0$, $\rho^{ions}_{0}=0$, $\rho^{ions}_{1}$, $r^{\prime}_{0}=\rho$
\For{$k = 1,...$}
\State $\rho^{tot}_{k} = r^{\prime}_{k-1} + \rho^{ions}_{k} - \rho^{ions}_{k-1}$
\State solve $\nabla \cdot \epsilon \nabla \phi^{\prime}_{k} = -4 \pi \rho^{tot}_{k}$ (Algorithm \ref{PCG} with a residual vector $r^{\prime}_{k}$)
\State $\phi_{k} = \phi_{k} + \phi^{\prime}_{k}$
\State compute $\rho^{ions}_{k+1} = \rho^{ions}[\phi_{k}]$
\State $\rho^{ions}_{k+1} = \eta \rho^{ions}_{k+1} + (1-\eta)\rho^{ions}_{k}$
\State $r_{k+1} = \rho^{ions}_{k+1} - \rho^{ions}_{k}$
\EndFor
\end{algorithmic}
\end{algorithm}
\end{figure}

\subsection{Numerical performances}
To show performances and accuracy of the Poisson-Bolzmann solver,
both Algorithms \ref{SCPB} and \ref{SCPBimproved} have been applied to analytical test cases.
Following the strategy reported in Sec. \ref{Nume}, all the involved fields and functions,
i.e. the potential, the charge density and the operator have been discretized on the
orthorhombic three dimensional grid and same parameters have been used to set up
all analytical functions.
The electrostatic problem lies in a cavity where the great majority of the charge density is confined,
described by means of a dielectric error function $\epsilon(\textbf{r})$ (solid black line
of Fig. \ref{AnFunc}). A normalized Gaussian function has been chosen for the electrostatic
potential $\phi(\textbf{r})$ (red dash line of Fig. \ref{AnFunc}), whereas the charge
density $\rho(\textbf{r})$ has been derived from the chosen potential and dielectric function,
applying the Poisson-Boltzmann differential operator.

\begin{figure}
\includegraphics[width=0.5\textwidth]{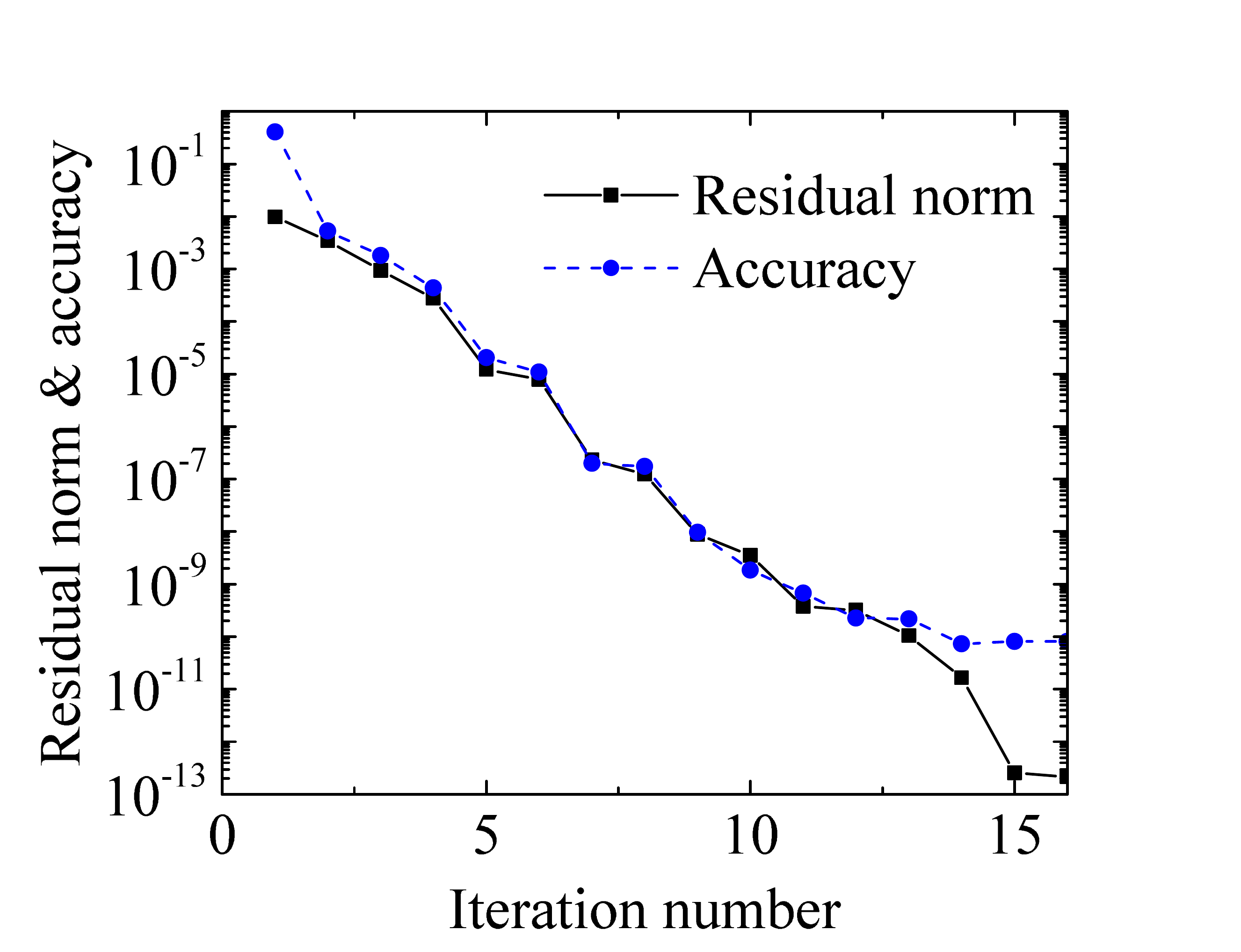}
\caption{\label{PBlinear_data}Euclidean norm of the residual
vector $r_k$ (black squares) and accuracy of the
numerical solution (blue circles) for the linear Poisson-Boltzmann
equation with free boundary conditions (Algorithm \ref{PCG}
has been applied with $\mathcal A$ the linear Poisson-Boltzmann operator).}
\end{figure}

A monovalent ($Z_1=-Z_2=1$) binary aqueous electrolyte solution
has been considered, with a close packing coefficient $p=0.74$, effective ionic radius
$R_{i}=3$ [\AA] and bulk ion concentrations $c_{i}^{\infty}=100$ [mol/m$^3$] kept fixed for all ions.
As discussed in Sec. \ref{PBequation}, the PCG algorithm
has been applied to solve the linear Poisson-Boltzmann equation
with the same GPe preconditioner of Eq. (\ref{PCGdef}).
In Fig. \ref{PBlinear_data} the euclidean norm of the residual vector
$r_k$ (black squares) and the accuracy of the numerical solution
(blue circles) have been reported. Similar performances
have been found with respect to the generalized Poisson solver,
reaching an accuracy of $\sim 10^{-10}$ with some ten iterations and
proving that the PCG procedure is a well suited and fast method
also for the linear regime of the PBe.

\begin{figure}
\includegraphics[width=0.5\textwidth]{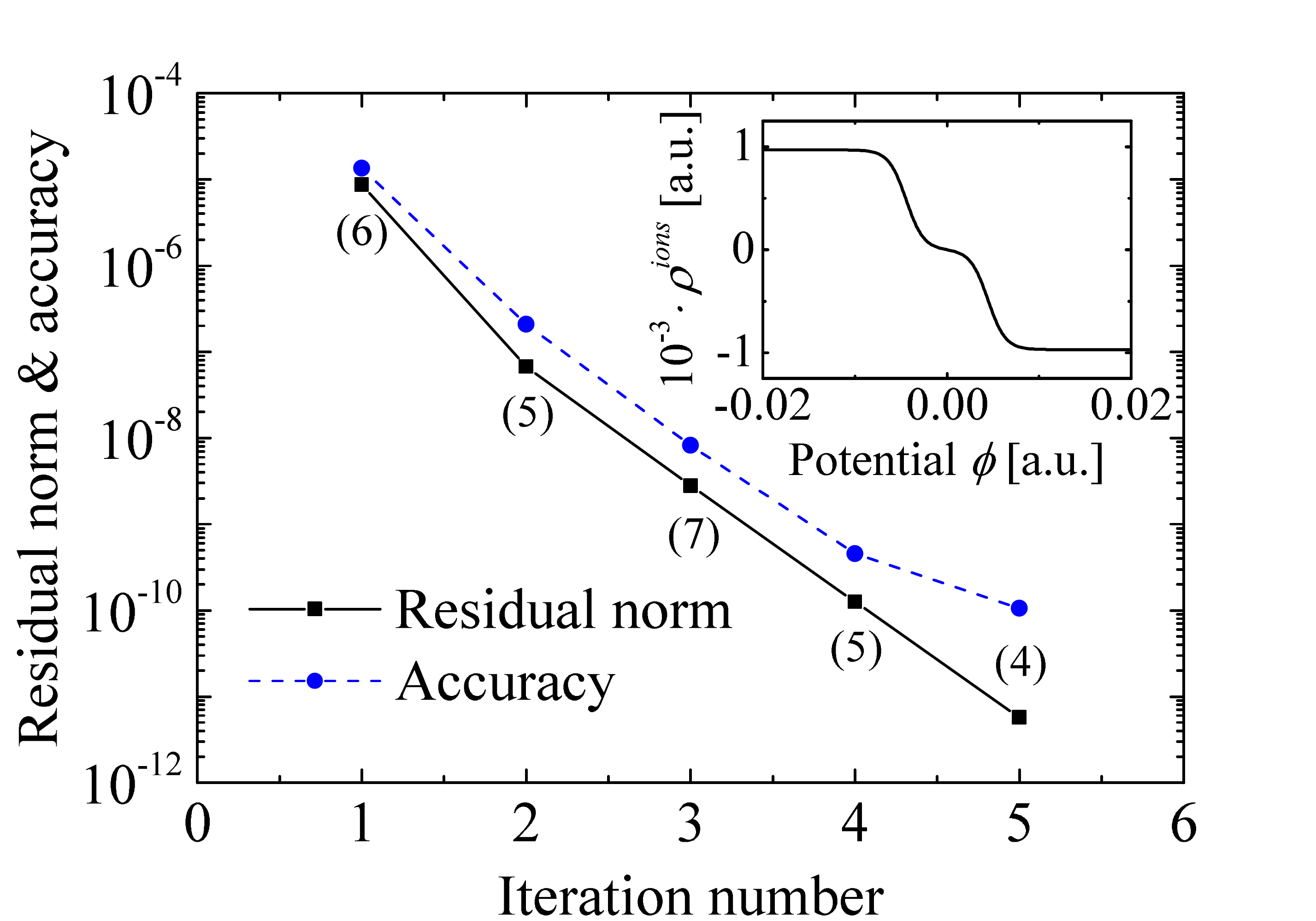}
\caption{\label{PB_data}Euclidean norm of the residual vector $r_k$ (black squares) and accuracy of the
numerical solution (blue circles) for the modified Poisson-Boltzmann
solver with free boundary conditions. Numbers between round brackets
represent the number of iterations needed to solve the GPe at step 4
of Algorithm \ref{SCPBimproved}.
The inset shows $\rho^{ions}[\phi]$ given by Eqs. (\ref{rhoions},\ref{ionsconcMPB}).}
\end{figure}

In the general non-linear case, Algorithms \ref{SCPB} and
\ref{SCPBimproved} have been applied.
An input guess $\rho^{ions}_{0}=0$ has been chosen.
We found that only a relatively small number of self-consistency iterations is needed in this approach
to solve both the Poisson-Boltzmann and the modified Poisson-Boltzmann problem.
Fig. \ref{PB_data} shows the Euclidean norm of the residual vector $r_k$ (black squares)
and the accuracy of the numerical solution (blue circles) for the modified Poisson-Boltzmann
solver with free boundary conditions.
The solver exhibits a fast convergence, reaching an accuracy of $\sim 10^{-10}$
with some five iterations.
The inset shows $\rho^{ions}[\phi]$ as given by Eqs. (\ref{rhoions},\ref{ionsconcMPB}),
revealing an ionic density saturation in the solvent for electrostatic potentials which absolute value is higher than $\sim 0.01$ [a.u.].
Numbers between round brackets
represent the number of iterations needed to solve the GPe at step 4
of Algorithm \ref{SCPBimproved} using the PCG scheme. The convergence criterion for the GPe solver has
been fixed equal to the one of Algorithm \ref{SCPBimproved},
except for the first two iterations where very accurate potentials
do not change the overall performances of the PBe solver.
Furthermore its behavior does not change with the boundary conditions
which are managed by means of the generalized Poisson solver.
The latter also fixes the final accuracy of the self-consistent procedure.
It is worth noting that performances and parallel efficiency as well as boundary conditions
of both GPe and (M)PBe solvers are, eventually, delegated to the underlying SPe solver.

\section{Electronic structure computations}
Effects of complex wet environments surrounding an atomistic system can be
approximately included into density functional calculations by simply
introducing a dielectric cavity surrounding the atomistic system and taking
in Eq. (\ref{ene}) the electrostatic potential $\phi[\rho(\textbf{r})]$
as solution of the generalized Poisson or Poisson-Boltzmann equation.
The PCG solver for the GPe (Algorithm \ref{PCG}) has been implemented in the electronic-structure
package BigDFT\cite{BigDFT}, extending the capability of the code
beyond vacuum-simulations.
Two distinct approaches have been implemented and tested to build up a dielectric cavity enveloping the atomic system.
In both approaches the cavity, mapped by the dielectric function $\epsilon(\textbf{r})$, is
fully differentiable and continuous in the whole simulation domain.

In the first approach a function $\epsilon(\textbf{r})$ is defined starting from
spherical-symmetric atom-centered cavities. Each sphere depends only
on the radial distance with respect to a fixed atomic position. The whole rigid cavity
is kept fixed during the whole SCF cycle in a DFT simulation.

On the other hand, it could be argued that regions occupied by atoms and their associated electronic charge density are
strictly related. In other words, the actual value of the charge density might determine
how much ``space'' is occupied by the solute. Starting from this idea, a cavity can be build up
directly from the electronic density, as it has been shown in various
publications~\cite{Fattebert_JCC2002,Scherlis_JCP2006,Andreussi_JCP2012}. In this second approach
the dielectric function is not explicitly space-dependent, but can be implicitly mapped by means of
the electronic charge density.

\subsection{\label{Rigcav} Rigid cavity}

The most widespread continuum solvation model, which tries to include the effects of a surrounding
dielectric medium in an implicit way, is the polarizable continuum model developed
by J. Tomasi and co-workers \cite{Miertus_CP1981,Tomasi_CR1994,Tomasi_CR2005}.
In the PCM formulation the cavity surrounding the solute is sharp and discontinuous, and a polarization charge density
is exactly localized at the vacuum-dielectric interface. In this way the dielectric environment is represented by
an effective surface polarization charge, reducing the electrostatic problem into a two-dimensional one.
Furthermore the cavity can be considered rigid since it depends only on the atomic coordinates
which does not vary during a SCF cycle in DFT simulations.

For a first test of the electrostatic solvers
(Algorithm \ref{SC} and \ref{PCG}) integration
in an electronic-structure code, a PCM-like cavity has been considered.
The dielectric function $\epsilon(\textbf{r})$ is given by the product of spherical-symmetric
atom-centered error functions.
Differently from the early PCM model, we here define a cavity which is fully differentiable and continuous.
In particular, for a system of $N$ atoms of coordinates $\textbf{R}_{i}$ (for $i = 1,...,N$)

\begin{equation}
\label{RigCav}
\epsilon( \textbf{r},\{\textbf{R}_{i}\}) = (\epsilon_0 -1) \left\{\prod_i h(d_i,\Delta;d(\textbf{r},\mathbf R_i))\right\} + 1 ,
\end{equation}

where $\epsilon_0$ is the dielectric constant of the surrounding medium and the function $h$ is defined by Eq.~\eqref{hfunc}.
In Eq. (\ref{RigCav}) $d(\textbf{r},\textbf{R}_{i})=\lVert \textbf{r} - \textbf{R}_{i} \rVert$, $d_{i}$
are the cavity radii which depend
on the particular atom species, and $\Delta$ a parameter (kept fix for all atoms) which controls
the transition
region ($\approx 4 \Delta$ wide) from 0 to 1 where the polarization charge is located.
Starting from Eq. (\ref{RigCav}), all vectors which explicitly depend on $\epsilon(\textbf{r})$
can be analytically computed. The cavity is uniquely defined once $\textbf{R}_{i}$,
$d_{i}$ and $\Delta$ are fixed.
All environment-dependent fields are calculated once at the start of the solver and kept fixed
throughout the procedure.

In must be noticed that this definition of the cavity relies on the explicit dependence
of $\epsilon(\textbf{r},\left\{\textbf{R}_{i}\right\})$ from the atomic
coordinates $\textbf{R}_{i}$  (and, consequently, of the system Hamiltonian).
This dependence gives rise to additional terms when atomic forces are computed.
The analytical rigid cavity above described should overcome this problem allowing a direct
analytic calculation of such additional contributions.
Moreover, the values of $d_i$ and of $\Delta$ have to be tuned by the user, usually by choosing a
solvent-dependent scaling factor with respect to empirical Van der Waals radii\cite{Orozco2000}.
A solution that would remove part of this arbitrariness should therefore avoid an explicit
use of atomic coordinates in the cavity mapping.
This will be the subject of the following section.

\subsection{\label{Sccav} Charge-dependent cavity}
For this definition of the cavity, the dielectric function does not explicitly depend in the
atomic positions, but implicitly via the charge density $\rho^{elec}$
\begin{equation}
\epsilon( \textbf{r}) = \epsilon[\rho^{elec}](\textbf{r}) .
\end{equation}

This approach allows the cavity to self-consistently change during the SCF loop, strictly following
the modification of the electronic charge density. 
A cavity surrounding an atomic system can
be generated by means of two threshold parameter, i.e. $\rho_{max}$ and $\rho_{min}$,
fixing $\epsilon( \textbf{r})=1$ in regions when $\rho^{elec}(\textbf{r}) > \rho_{max}$ and
$\epsilon( \textbf{r})=\epsilon_0$ when $\rho^{elec}(\textbf{r}) < \rho_{min}$.
Like $d_{i}$ in the rigid case, $\rho_{max}$ fix the width of the cavity, whilst the extension
of the transition region, previously defined by $\Delta$, is now tuned by $\rho_{min}$.

Several features make the self-consistent cavity more advantageous with respect to
the rigid one. First, once the electron charge density
is given, only two parameters uniquely define both the cavity and
the transition region for the whole atomic system.
Furthermore, since the dielectric function does not explicitly depend on the atomic coordinates, the evaluation of ionic forces can be done without modifications with respect to a simulation in gas-phase.

Among several ways to define the functional dependence on the electronic charge density,
the self-consistent continuum solvation (sccs) model developed in Ref. [\citenum{Andreussi_JCP2012}] has been implemented.
It allows to fit experimental solvation energies on
a set of 240 neutral solutes with a mean absolute error of 1.2 kcal/mol:

\begin{equation}
\linespread{1.5}\selectfont
\label{sccs}
\epsilon( \rho^{elec}) = \begin{cases} 1 & \rho^{elec} > \rho_{max} \\
\displaystyle \epsilon_0 e^{w(\rho^{elec})} & \rho_{min} < \rho^{elec} < \rho_{max}\\
\epsilon_0 & \rho^{elec} < \rho_{min}\end{cases} ,
\end{equation}

where $w(x)$ is a continuous smooth function describing the transition region between vacuum (where atoms are placed) and
the full dielectric medium:
\begin{align}
w(x) &= \frac{1}{2 \pi} \left[ z(x) -\sin \left( z(x) \right) \right]; \\
z(x) &= 2 \pi \frac{\ln\left(\frac{\rho_{max}}{x}\right)}{\ln \left(\frac{\rho_{max}}{\rho_{min}}\right)}\;.
\end{align}

A good description of this region by means of Eq. (\ref{sccs}) is mandatory for the procedure
convergence as well as for the mapping of the polarization charge there confined.
Since $\epsilon(\textbf{r})$ explicitly depends on $\rho^{elec}(\textbf{r})$, its variation needs to be
included in the Kohn-Sham (KS) potential.
Starting from Eq. (\ref{GPe}) and integrating by parts, the electrostatic energy can be rewritten as

\begin{equation}
E_{es}[\rho] = \frac{1}{2} \int \rho \phi[\rho] \mathrm{d}\textbf{r}  = \frac{1}{8 \pi} \int \epsilon[\rho] (\nabla \phi[\rho])^2 \mathrm{d}\textbf{r} .
\end{equation}

Its functional derivative with respect to $\rho$ gives the electrostatic potential $\phi$ plus an 
additional term $v_{\epsilon}( \textbf{r})$
\begin{equation}
\label{veps}
v_{\epsilon}( \textbf{r}) = -\frac{1}{8 \pi} \frac{\mathrm{d}\epsilon(\rho^{elec}(\textbf{r}))}{\mathrm{d}\rho^{elec}} | \nabla \phi( \textbf{r})|^{2}\;,
\end{equation}
which has to be added to the KS potential.

\subsection{Solvation Free energies}

In order to test the integration and performances of the generalized Poisson solver
into \emph{ab initio} codes, the whole procedure previously described,
i.e. Algorithm \ref{SC} and \ref{PCG} of Sec. \ref{SCprocedure}
and \ref{PCGprocedure}, rigid and charge-dependent cavities of Sec. \ref{Rigcav} and \ref{Sccav}
as well as the additional term of Eq. (\ref{veps}) have been integrated in the main BigDFT
package \cite{BigDFT}.
This extension allows to handle complex wet environments in electronic-structure calculations,
including implicitly the effects of a solvent surrounding an atomic system.

The electrostatic solvation energy is defined as difference between the total energy
of a given atomic system in the presence of the dielectric environment and the energy
of the same system in vacuum

\begin{equation}
\label{Gel}
\Delta G^{el} = G^{el} - G^{0} .
\end{equation}

A full comparison with experimental solvation energies needs the inclusion
of non-electrostatic contributions.
In this case the main terms in the solute Hamiltonian, as introduced by PCM \cite{Tomasi_CR2005}, are

\begin{equation}
\Delta G^{sol} = \Delta G^{el} + G^{cav} + G^{rep} + G^{dis} + G^{tm} + P \Delta V ,
\end{equation}

where $\Delta G^{el}$ is the electrostatic contribution, $G^{cav}$ the cavitation energy,
i.e. the energy necessary to build up the solute cavity inside the solvent medium.
$G^{rep}$ is a repulsion term representing the continuum counterpart
of the short-range interactions induced by the Pauli exclusion principle,
whilst $G^{dis}$ reflects van der Waals interactions.
The thermal term $G^{tm}$ accounts for the vibrational and rotational changes and, finally, $P \Delta V$
includes volume change in the solute Hamiltonian.

The inclusion of all non-electrostatic contributions goes beyond the aim of the present paper,
where testing of the generalized Poisson solver and its integration into first principle
atomistic calculations is the main goal.
Therefore only electrostatic solvation energies have been computed for a set of
small neutral organic molecules.

Water has been chosen as solvent for all cases, with a dielectric constant
of $\epsilon_0 = 78.36$ (experimental value at low frequency and ambient conditions).
Final energies for all molecules have been extracted after a full geometry optimization
both in vacuum and in aqueous solution.
In all cases the surrounding dielectric medium lowers the total energy of the system
with respect to vacuum, because the polarization of the dielectric stabilizes the solutes.

Table \ref{TabEsol} reports electrostatic solvation energy $\Delta G^{el}$
obtained both with rigid and self-consistent cavities under free boundary conditions.
As previously stated, a critical point of PCM approaches is the choice of shape and size
of the cavity, which should mimic the solute incorporating the whole atomic charge density.
From its first formulation \cite{Miertus_CP1981}, PCM atomic radii $R_{i}$ were
fixed proportional to the van der Waals radii, $R_{i}=f \cdot R_i^{vdW}$.
In our rigid model, a proportional factor of $f=1.2$ has been fixed and the Pauling's set of atomic radii has been
considered \cite{Mineva_JCC1998,Pauling_Book1960,Hbcp_web} (except for the hydrogen atoms bound
to heteroatoms which have a radius value of 1 \AA ).
Having a further degree of freedom with respect to sharp PCM cavities, a $\Delta=0.5$ [a.u.] has
been tuned and kept constant for all atoms.

As reference, in Table \ref{TabEsol} sccs calculations from Ref. [\citenum{Andreussi_JCP2012}]
together with PCM calculations have been reported. 
The latter were performed with the {\sc gaussian} 09 code\cite{g09}, but using the version
of PCM which was the default in {\sc gaussian} 03, as specified by the keyword g03defaults.
Only electrostatic solvation effects were included in the calculation and, to simplify the comparison,
the simple van der Waals surface was adopted with Pauling's set of atomic radii\cite{Pauling_Book1960}
(explicit hydrogens), but without the additional smoothing used to describe the solvent-excluded surface.
The Perdew-Berke-Ernzerhof (PBE) functional \cite{Perdew_PRL1996} and the extended triple-zeta 6-311+g(d,p) basis set were used for both geometry optimizations and energy calculations, in vacuum and in solution, consistently with the set up used for the
parameterization of the electrostatic solvation energy in sccs\cite{Andreussi_JCP2012}. 

\begin{table}
\caption{\label{TabEsol}Electrostatic solvation energies $\Delta G^{el}$ (in kcal/mol)}
\begin{ruledtabular}
\begin{tabular}{lllll}
 & PCM\footnotemark[1] & rigid$_{\text{BigDFT}}$ & sccs$_{\text{QE}}^P\footnotemark[2]$ & sccs$_{\text{BigDFT}}^F$\\
\hline
 NH$_{3}$           &  -6.65  &  -6.28 &  -5.39 &  -5.35\\
 H$_{2}$O           &  -8.98  &  -8.32 &  -8.21 &  -8.23\\
 CH$_{4}$           &  -0.61  &  -1.20 &  -0.68 &  -0.63\\
 CH$_{3}$OH         &  -6.78  &  -6.57 &  -5.89 &  -5.83\\
 CH$_{3}$NH$_{2}$   &  -4.51  &  -5.71 &  -4.53 &  -4.45\\
 CH$_{3}$CONH$_{2}$ & -12.53  & -12.97 & -11.87 & -11.87\\
\end{tabular}
\end{ruledtabular}
\footnotetext[1]{Polarizable continuum model results obtained with {\sc gaussian} 09\cite{g09} and Pauling's set of atomic radii\cite{Pauling_Book1960}}
\footnotetext[2]{Self-consistent continuum solvation model results from periodic BC calculations in Ref. \cite{Andreussi_JCP2012}}
\end{table}

In all calculations soft norm-conserving pseudopotentials including non-linear core correction \cite{Goedecker_PRB1996,Willand_JCP2013}
along with PBE functional were
used to describe the core electrons and exchange-correlation, respectively.

For the self-consistent cavity, 
values of $\rho_{max} = 5\cdot 10^{-3}$ and $\rho_{min} = 1\cdot 10^{-4}$ have been fixed which
produces a mean absolute error of 1.20 kcal/mol for the solvation energies of a database
of 240 molecules\cite{Andreussi_JCP2012}.
A perfect agreement has been reached which confirms the performance
and reliability of the integrated generalized Poisson solver.

In order to test and validate the whole library in BigDFT, the effects of grid resolution
and boundary conditions on the electrostatic solvation energies for the same set of neutral
molecules have been investigated.
The global accuracy is strictly related to the size of the simulation box and the spatial
grid resolution $h_{\text{grid}}$. However a decrease of the latter can affect the
whole cost of the calculation both in time and memory usage.
Consequently it is worth to investigate the effects of the dielectric
medium's inclusion in a DFT run with respect to the vacuum case.

\begin{figure}[b]
\includegraphics[width=0.5\textwidth]{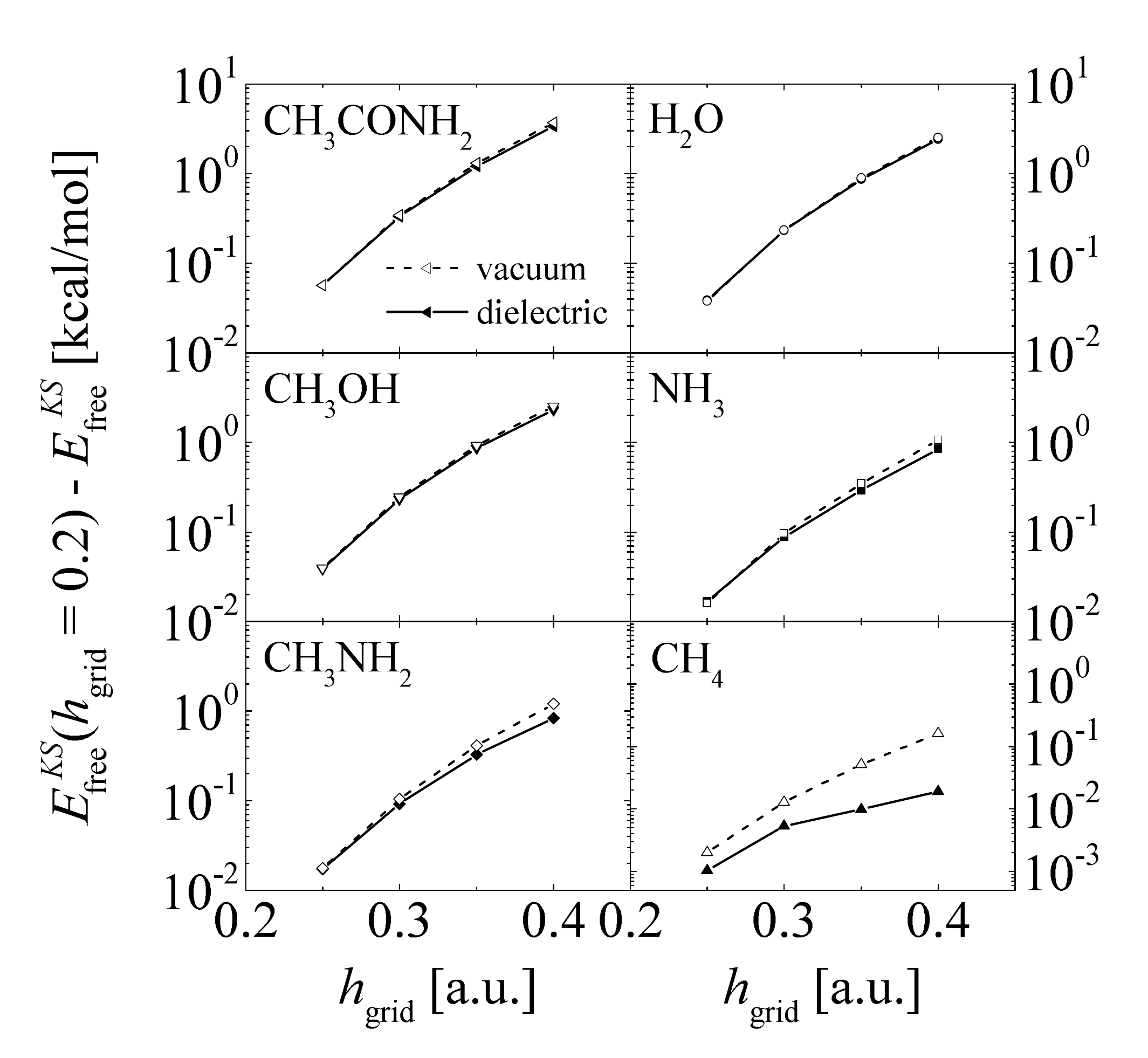}
\caption{\label{EKS_diff}Accuracy of the Kohn-Sham total energy
with respect to the spatial grid $h_{\text{grid}}$.
Calculations for all molecules have been done
both in vacuum (empty symbols, dash lines) and
with a surrounding dielectric environment (filled symbols, solid lines)
in free boundary conditions.}
\end{figure}

\begin{figure}
\includegraphics[width=0.5\textwidth]{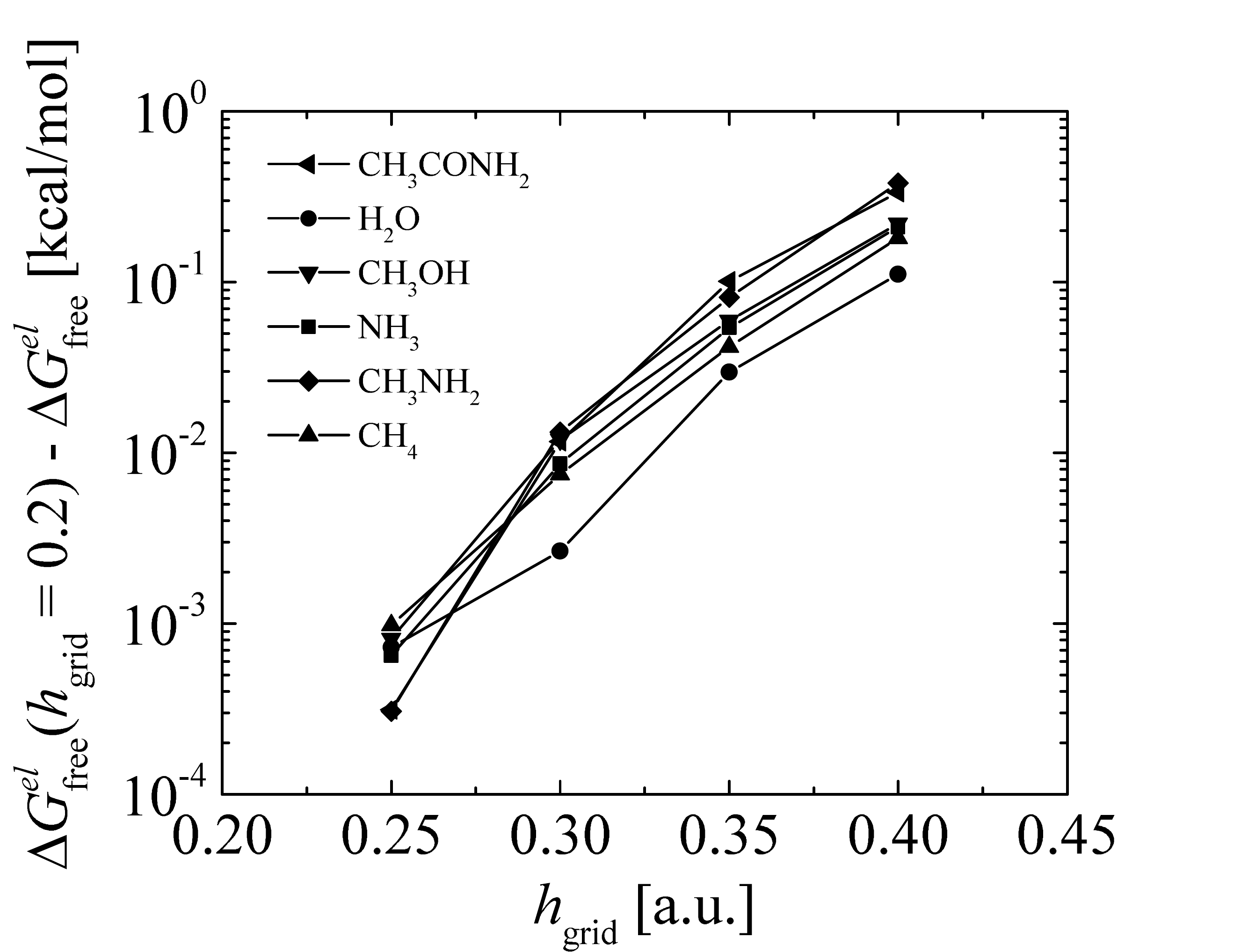}
\caption{\label{Hgrid_sol}Accuracy of the electrostatic
solvation energy with respect to the spatial grid $h_{\text{grid}}$
with free boundary conditions.}
\end{figure}

In this respect Kohn-Sham total energies have been extracted both in vacuum and in the presence of the
solvent (water) as function of the spatial grid $h_{\text{grid}}$. Its accuracy is reported in
Fig. \ref{EKS_diff} as difference with the reference value at $h_{\text{grid}} = 0.20$ bohr.
Results show that the accuracy is not affected by the surrounding dielectric environment
(filled symbols, solid lines) with respect to the vacuum case (empty symbols, dash lines).
Following the same guidelines,
the accuracy of the electrostatic solvation energy [Eq. (\ref{Gel})] with respect to the spatial grid $h_{\text{grid}}$
has been investigated and reported in Fig. \ref{Hgrid_sol}.
Results show that a $h_{\text{grid}} = 0.30$ bohr in BigDFT with free boundary condition
provides electrostatic solvation energies with an accuracy lower than $\sim 10^{-2}$ kcal/mol.

\begin{figure}[b]
\includegraphics[width=0.5\textwidth]{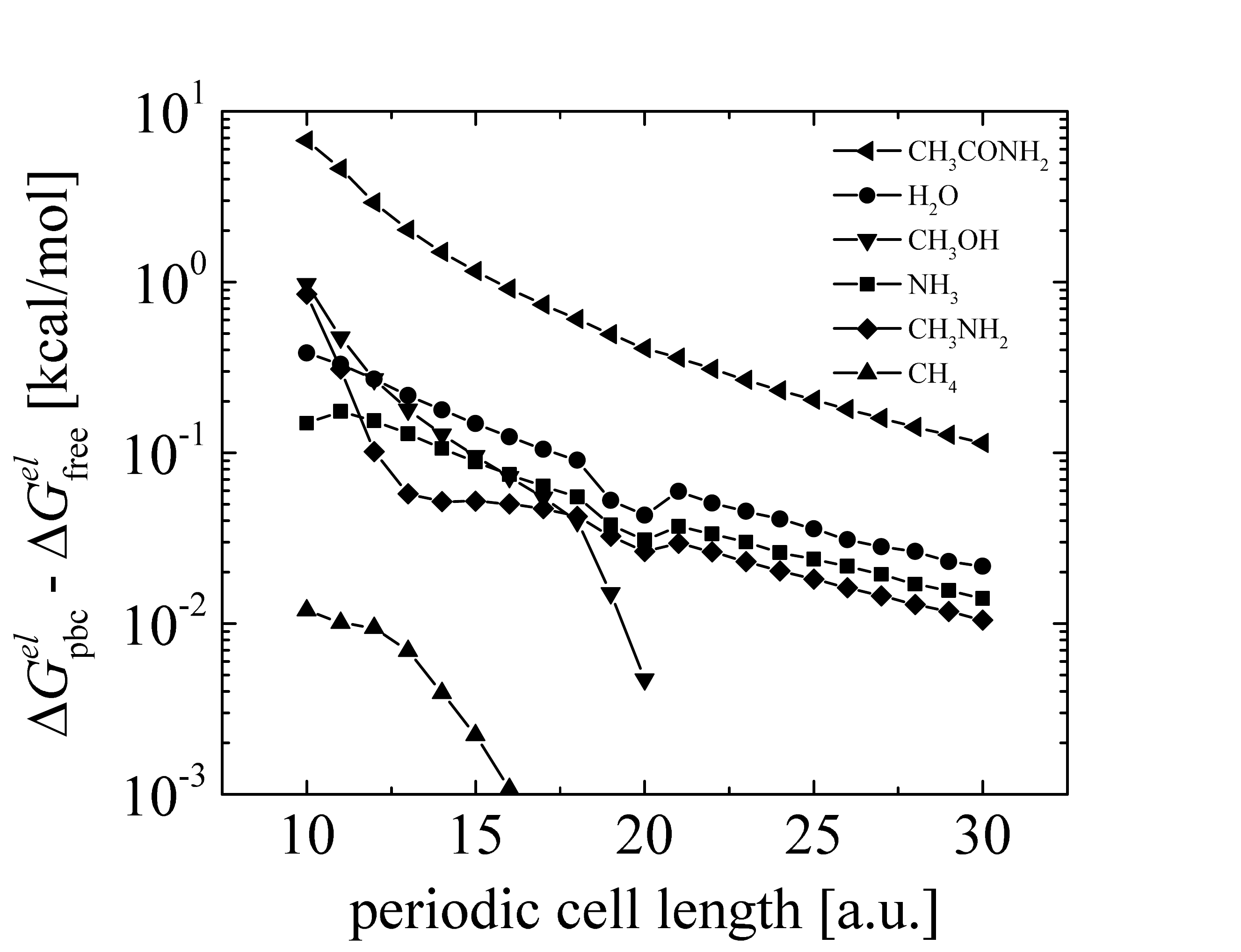}
\caption{\label{Hcell_sol}Difference between the electrostatic
solvation energy computed with periodic and free boundary
conditions as function of the periodic cell length.
Molecules have been ordered as function of their electrostatic dipole norm, from largest
to smallest (see Table \ref{TabDipole}).}
\end{figure}

Since isolated systems embedded in a wet environment are the subject of interest, spurious
long range electrostatic interactions with periodic images due to artificially imposed 
periodicities along certain direction can be problematic.
In the developed procedure the boundary conditions enter through the preconditioner, i.e. the SPe solver.
Since in the BigDFT Poisson solver all the common boundary conditions such as free, wire and surface
are exactly implemented, such spurious interactions do not exist at any stage of our approach.
Fig. \ref{Hcell_sol} reports the difference between the electrostatic solvation energies computed
with periodic and free boundary conditions as function of the periodic cell length and
keeping fixed the spatial grid $h_{\text{grid}} = 0.20$ bohr. 
Molecules have been ordered according to the strength of their electrostatic dipole, from largest to smallest. 
Dipole values for each molecule are reported on Table \ref{TabDipole} both in vacuum and in the present of
a water solvent. It can be noticed that the presence of the polar solvent increases the electrostatic dipoles 
of the molecules~\cite{Orozco2000}.
As it might be expected, the free BC calculation represents the asymptotic results for periodic
BC calculation of increasing box sizes.
Interactions with image-molecules are less relevant for
molecules with small dipole moment like CH$_{4}$, but are not negligible for molecules with
larger dipoles such as CH$_{3}$CONH$_{2}$.
In such cases unrealistic large periodic boxes are required for periodic BC to reproduce the free boundary
condition results with high accuracy. 

\begin{table}
\caption{\label{TabDipole}Molecule dipole norm in vacuum and water solvent (in Debye).}
\begin{ruledtabular}
\begin{tabular}{lll}
 & Dipole$^{vacuum}$ & Dipole$^{water}$\\
\hline
 CH$_{3}$CONH$_{2}$ & 3.88 &  5.76\\
 H$_{2}$O           & 1.81 &  2.41\\
 CH$_{3}$OH         & 1.57 & 2.14\\
 NH$_{3}$           & 1.49 & 1.98\\
 CH$_{3}$NH$_{2}$   & 1.27 & 1.78\\
 CH$_{4}$           & 0.00 & 0.00\\
\end{tabular}
\end{ruledtabular}
\end{table}

\begin{figure}
\includegraphics[width=0.45\textwidth]{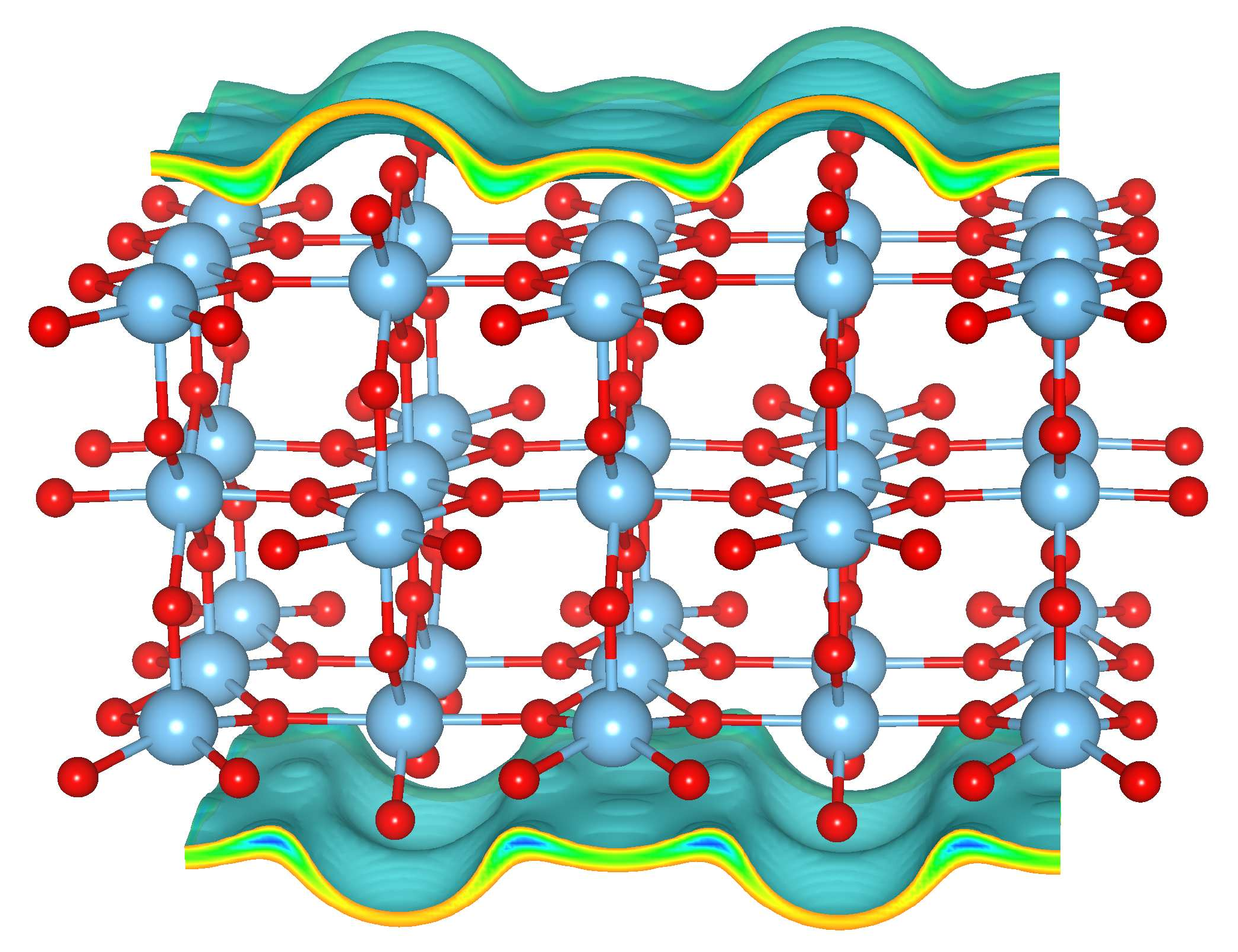}
\caption{\label{TiO2} TiO$_2$ surface in contact with water: isosurfaces
of the polarization charge $\rho^{pol}$ in the implicit dielectric medium.}
\end{figure}

Numerous processes of practical interest involve surfaces in contact with neutral or ionic solvents,
leading to an induced  polarization charge of the dielectric medium or an electric
double layer. The BigDFT package allows to use exact surface boundary conditions
avoiding spurious interaction in the direction orthogonal to the surface exposed
to the wet environments. To show a further application of the solvation library
in BigDFT, a TiO$_2$ surface in contact with pure water has been simulated.
The full DFT simulation in presence of the solvent has been initialized
starting from its relaxed state in vacuum. 
Fig. \ref{TiO2} shows the TiO$_2$ wet surface as well as isosurfaces of the
polarization charge density $\rho^{pol}$ as given by Eq.~\eqref{rhopol}. 
In this test case, we found that
the contact with the dielectric medium provides an electrostatic solvation energy of -46.60 kcal/mol.

\section{Conclusions}
In the present work a library to handle complex wet-environments in electronic-structure calculations
has been presented. It allows to include on the atomistic scale effects of an aqueous environment
in an implicit way. The solver is able to handle both the generalized Poisson
and several variants of the Poisson-Boltzmann equation.
The core of the generalized Poisson solver is a preconditioned conjugate gradient
algorithm which allows to numerically solve the minimization problem with some ten iterations.
The same algorithm works for the linear case of the Poisson-Boltzmann equation, whilst for the general
case a self-consistent procedure has been implemented.
The chosen preconditioner is based on the ISF Poisson solver for the standard
Poisson equation, which can handle all common boundary conditions exactly.
The code requires a small amount of memory and is very fast and in addition 
also highly parallelized. 
We have shown that coupled with BigDFT, our method can correctly reproduce
electrostatic solvation energies of a set of small neutral organic molecules.
Effects of grid resolution and boundary conditions on the electrostatic solvation energies
have been reported.
The whole library, will be released as an independent program 
suitable for integration in other electronic structure codes.
A GPU-accelerated version of this software package is also in preparation, following the guidelines indicated in Ref. [\citenum{dugan2013}].

\begin{acknowledgments}
This work was done within the PASC and NCCR MARVEL projects. Computer resources were provided
by the Swiss National Supercomputing Centre (CSCS) under Project ID s499. LG acknowledges also support from the EXTMOS EU project.
\end{acknowledgments}

\bibliography{Reference}

\end{document}